\documentclass[twocolumn,showpacs,prb,floatfix,superscriptaddress,citeautoscript
,
cite ] { revtex4}
\usepackage{graphicx}
\usepackage{color}

\begin{document}

	\author{H. Sahin}
		\email{hasan.sahin@ua.ac.be}
		\affiliation{Department of Physics, University of Antwerp,
Groenenborgerlaan 171, B-2020 Antwerpen, Belgium}

	\author{J. Sivek}
		\affiliation{Department of Physics, University of Antwerp,
Groenenborgerlaan 171, B-2020 Antwerpen, Belgium}

	\author{S. Li}
		\affiliation{Department of Physics, University of Antwerp,
Groenenborgerlaan 171, B-2020 Antwerpen, Belgium}
		\affiliation{Nano Structural Materials Center, Nanjing
University of Science and Technology, Nanjing, China}

	\author{B. Partoens}
			\affiliation{Department of Physics, University of
Antwerp, Groenenborgerlaan 171, B-2020 Antwerpen, Belgium}

	\author{F. M. Peeters}
		\affiliation{Department of Physics, University of Antwerp,
Groenenborgerlaan 171, B-2020 Antwerpen, Belgium}

\title{Stone-Wales Defects in Silicene: Formation, Stability and Reactivity of
Defect Sites}

\date{\today}


\begin{abstract}

During the synthesis of ultra-thin materials with hexagonal lattice
structure Stone-Wales (SW) type of defects are quite likely to be formed and the
existence of such topological defects in the graphene-like structures results in
dramatical changes of their electronic and mechanical properties. Here we
investigate the formation and reactivity of such SW defects in silicene. We
report the energy barrier for the formation of SW defects in freestanding
($\sim$2.4~eV) and Ag(111)-supported ($\sim$2.8~eV) silicene and found it to be
significantly lower than in graphene ($\sim$9.2~eV). Moreover, the buckled
nature of silicene provides a large energy barrier for the healing of the SW
defect and therefore defective silicene is stable even at high temperatures.
Silicene with SW defects is semiconducting with a direct bandgap of 0.02 eV and
this value depends on the concentration of defects. Furthermore, nitrogen
substitution in SW defected silicene shows that the defect lattice sites are the
least preferable substitution locations for the N atoms. Our findings show the
easy formation of SW defects in silicene and also provide a guideline for
bandgap engineering in silicene-based materials through such defects.

\end{abstract}

\maketitle

\section{Introduction}

The electronic, thermal and mechanical properties of graphene\cite{novo,
moriz,chen,balan,lee,fric} are sensitive to lattice imperfections and adatoms.
However, the creation of in-plane defects, such as vacancies, adsorption and
doping of foreign atoms on the honeycomb structure, opens the possibility for
tailoring the electronic and magnetic properties of graphene-based
structures.\cite{Banhart,balog, batzill, Tapa, Kras, gass, coat} It was
demonstrated
that Stone-Wales (SW) defects\cite{stone} can be formed during a rapid quench
from high temperature or when graphene is under irradiation.\cite{Meyer}
SW defects have been predicted to open a band gap in the electronic
band structure that can be of potential use in the design
of transistors.\cite{sw1,sw2,sw3,Lusk} Moreover presence of such defects modify
the chemical reactivity of the graphene lattice.\cite{li,chen2} Very recently,
Ij\"{a}s \textit{et al}. have reported that the presence of SW defects in honeycomb
lattice results in more reactive domains for chlorine atoms as compared to
defect-free graphitic structures.\cite{mari}

Nowadays, active research on ultra-thin materials has also been directed towards
the synthesis and manipulation of monolayer silicon called silicene. In
a silicene structure a hexagonal mesh of silicon atoms is buckled and studies
show that its electronic properties are similar to those of
graphene.\cite{takeda, cahan, hasan} After early theoretical predictions very
recent experimental studies have revealed clear evidence of the existence of
monolayer honeycomb structures of silicene. In recent works of Vogt \textit{et
al.}\cite{vogt} and Feng \textit{et al.}\cite{Feng} high-quality large-scale
silicene films have been synthesized on the Ag (111) surface. Moreover,
Fleurence \textit{et al.} demonstrated the successful growth of epitaxial
silicene on a diboride surface.\cite{Fleurence} Recent theoretical studies have
also reported that, prior to graphene, silicene has unique features such as a
large spin-orbit gap at the Dirac point,\cite{silicene-soc} experimentally
accessible quantum spin Hall effect,\cite{silicene-qshe} electrically tunable
band gap\cite{silicene-falko} and the emergence of valley-polarized metal
phase.\cite{motohiko} Most recently we reported the effect of impurities on the
structural, electronic, magnetic properties and lattice dynamics of
silicene.\cite{silicene-Hasan,silicene-Jozef}

Although the SW type defects are quite likely to be formed in such monolayer
systems, the formation of SW-type defects and their effect on the electronic,
magnetic and adsorption characteristics of monolayer silicene have remained
unexplored. In this paper, we use first-principles calculations within the
density functional theory (DFT) formalism to investigate the formation and the
electronic properties of SW defects in silicene. We found that formation SW
type defects are more likely in silicene than in graphene and the existence of
the underlying Ag(111) surface does not significantly affect their formation.
Furthermore, once the SW defect is formed in the silicene lattice the doping
characteristics are influenced dramatically.

This paper is organized as follows. Section \ref{methodology} describes the
computational methodology. Section \ref{results}-A discusses the formation
and stability of SW defects in silicene. Section \ref{results}-B presents the
energetics of N-doping in SW-defected and defect-free silicene. Findings are
concluded in Section \ref{conclusion}.

\section{Computational Methodology}\label{methodology}

In the present work, we performed first-principles calculations based on the
plane-wave basis set with a cut-off energy of 500 eV and the projector-augmented
wave (PAW) pseudopotentials implemented in the Vienna Ab-initio Simulation
Package (VASP).\cite{kresse1, kresse2} The exchange-correlation functionals are
described by the generalized gradient approximation (GGA) with
Perdew-Burke-Ernzerhof (PBE) approach.\cite{perdew} The partial occupancies for
the total energy ground state calculations were calculated with the tetrahedron
methodology with Bl\"ochl corrections \cite{blochl_1994-tetrahedron}.

Periodic boundary conditions were employed for silicene with a vacuum region of
15~\AA~between neighboring slabs. For supported silicene on a Ag (111)
surface the height of the supercell has been set to 29~\AA~in order to
conveniently include 4 layers of the Ag crystal. In order to determine the
equilibrium configuration of silicene with defects, we relaxed all the atomic
coordinates and the supercell geometry using the conjugate gradient (CG)
algorithm with the maximum residual force of less than 0.01 eV/\AA. The sampling
of the
Brillouin zone was done for the supercell with the equivalent of
24$\times$24$\times$1 Monkhorst-Pack \cite{monk} k-point mesh for a silicene
unit cell composed of two Si atoms. In order to minimize the interactions
between the neighboring SW defects, for the calculation of geometric
and electronic properties, we employed a large supercell derived from a
6$\times$6 supercell of silicene where the distance between neighboring defects
is at least 14.6~\AA. The calculations of the formation energetics of a SW
defect for supported silicene were performed with a 6$\times$6 supercell of
silicene on top of 8$\times$8 supercell of Ag (111) surface.

\begin{table}
	\caption{\label{table1}
	Calculated structural and electronic properties for silicene and
	SW-defected silicene: lattice parameter ($a$), Si-Si distance ($d_{Si-Si}$),
	angle between two of the lattice vectors of a 6$\times$6 supercell ($\alpha$), total
	magnetic moment of the system ($\protect\mu_{Tot}$), formation energy ($E_{f}$),
	cohesive energy ($E_{coh}$) and the band gap ($E_{gap}$).}

	\begin{tabular}{ccccccccccccc}
	
	\hline\hline
                        &  $a$  & $d_{Si-Si}$ & $\alpha$ & $\mu _{Tot}$ & $E_{f}$ & $E_{coh}$ &$E_{gap}$\\
                        & \AA   & \AA         & deg      & $(\mu _{B})$ & $(eV)$ &   $(eV)$   & $(meV)$ \\
	\hline
	\textbf{Silicene}   & 23.21 & 2.28        & 60.0  & 0 & 0.00 & -3.98 &
1.5 \\
	\textbf{SW-Silicene}& 23.04 & 2.19-2.28   & 61.5  & 0 & 1.64 & -3.96 & 20 \\

	\hline\hline
	\end{tabular}
\end{table}

\begin{figure}
	\includegraphics[width=8.5cm]{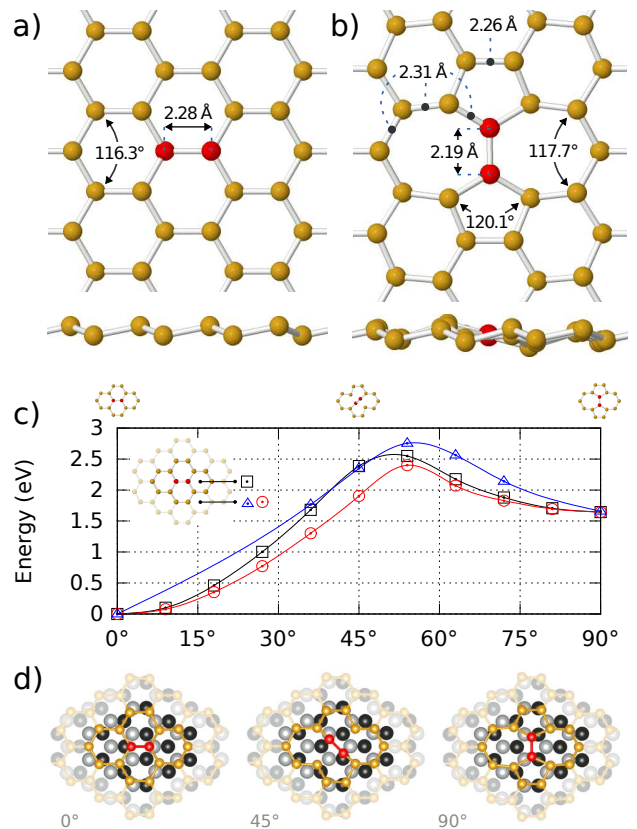}
	\caption{\label{fig-silicene_and_sw}
	(Color online) The top and side view of (a) perfect and (b) SW-defected
	silicene. (c) Energetics of formation of a SW defect in silicene via
	the bond rotation (highlighted). The black (open squares) curve is
	obtained by relaxing the first nearest hexagons to the silicene dimer in
	free-standing silicene layer. Red (open circles) and blue (open triangles)
	curves are obtained by relaxing up to the second nearest hexagons to the silicon
	dimer in the free-standing and supported silicene (on a Ag (111) surface),
	respectively. (d) The actual bond rotation steps (0${^\circ}$, 45${^\circ}$ and
	90${^\circ}$) in silicene supported on the Ag (111) surface.
	The uppermost and lower-lying Ag atoms are shown by gray (light) and black (dark).}

\end{figure}

\section{Results and Discussion}\label{results}

\subsection{Formation and Stability of Stone-Wales Defects in Silicene}

Since the formation of defects is inevitable in crystals, understanding their
effect on the mechanical, electronic and structural characteristics plays a key
role in nanoscale device applications. The most common defects that have been
observed in low-dimensional graphene-like structures are missing atoms, adatom
impurities and SW type of disorders. Here we investigate the formation and
stability of SW defects in a single layer structure of silicene shown in Fig.
\ref{fig-silicene_and_sw}(a). Compared to the graphitic materials such as
graphite, graphene and carbon nanotubes, the inter-atomic distance is larger in
silicene. Furthermore, due to the buckled hexagonal lattice structure, easier
formation of various defects in silicene can be expected. 

As shown in Fig. \ref{fig-silicene_and_sw}(b) a SW defect can be created by
the rotation of a silicon dimer by 90${^\circ}$ around the center of the Si-Si
bond. From the experimental point of view, such a defect can be formed during
the growth process or upon application of irradiation. After the formation of
the SW defect, four neighboring hexagons of silicene are transformed into a
pentagon and a heptagon pair. As can be seen from Fig.
\ref{fig-silicene_and_sw}(b), even after the creation of the SW defect, silicene
maintains its buckled two-dimensional structure with only a local out-of-plane
displacement of the Si atoms. Conversely, large local bumps in the graphene
lattice are formed due to the existence of SW defects. Through 90${^\circ}$
rotation of a dimer, the Si-Si bond becomes stronger than in defect-free
silicene and its length decreases from 2.28 to 2.19~\AA. After the shortening of
the bond lengths along the direction parallel to the pentagons, the lattice
constant decreases from 23.21 to 23.04~\AA. Though Si atoms favor buckled
configuration in bare silicene the formation of unbuckled
Si-dimer in the core of the SW-defect is preferable. Structural properties,
energetics and energy band gap of bare and SW-defected silicene are given in
Table \ref{table1}.

\begin{figure}
	\includegraphics[width=8.5cm]{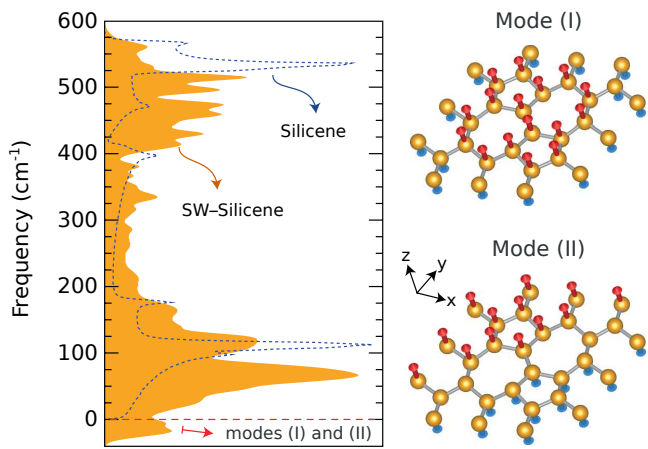}
	\caption{\label{fig-sw-phonon}
	(Color online) Phonon dispersion of pure (dashed line) and SW-defected silicene (filled curve).
	Atomic motions for two phonon modes which have imaginary eigenfrequencies
	are shown on the right. Only the motions of atoms in defect core and
	neighboring atoms are shown for clarity. Blue and red arrows represent
	displacements of Si atoms along $-z$ and $+z$ directions, respectively.}

\end{figure}

The transition states and energetics of the formation steps of the SW defect
in free standing silicene as well as for silicene supported on the Ag (111)
surface are shown in Fig. \ref{fig-silicene_and_sw}(c). The intermediate steps
are obtained by confining Si atoms at positions interpolated between those of
perfect silicene and silicene with a SW defect. The atoms that form two
heptagons and two pentagons of the SW defect and the nearest hexagons are
allowed to relax. The Si dimer that is gradually rotated is allowed to relax
only in the direction perpendicular to the Si layer. It is seen that the
formation of such SW defect in a perfect silicene lattice may occur by
overcoming the energy barrier of $\sim$2.4~eV for free standing silicene and
2.8~eV for the supported silicene layer. The maximum of the energy barrier
corresponds to the $\sim$54${^\circ}$ rotation of the bond. In comparison, this
barrier in graphitic materials is considerably larger and amounts to 9.2
eV.\cite{li} However, the energy barrier of 0.76 eV (1.10 eV for supported
silicene) between the intermediate configuration and silicene with a SW defect
guaranties the stability of such defects even at high temperatures. This shows
that the buckled lattice structure and the softer bonding nature of silicene
allows an easier formation of SW-defected structure. The formation energy,
$E_{f}$, of a SW defect in silicene is given by

\begin{equation}
	E_{f} = E_{Silicene}^{SW} - E_{Silicene}
\end{equation} 

where $E_{Silicene}^{SW}$ is the total energy of silicene with SW defect
and $E_{Silicene}$ is the total energy of perfect silicene. According to this,
the formation energy of a SW defect in a 6$\times$6$\times$1 supercell is
calculated to be 1.64 eV. Additionally, the stability of the structure can also
be predicted via its cohesive energy which is defined as the energy required
for separating the crystal into isolated free atoms. 

The cohesive energy of a crystal structure is given by the formula 
\begin{equation}
	E_{coh} = (E_{Tot}/n) - E_{Si}
\end{equation} 

where $E_{Tot}$ is the total energy of the system, $n$ is the number of atoms in
the supercell and $E_{Si}$ is the total energy of an isolated silicon atom. The
cohesive energies of silicene and SW-defected silicene are calculated to be
$-3.98$ and $-3.96$~eV, respectively. The negative cohesive energies of both
structures indicate their stability.

Next we extend our discussion on the stability and formation of SW-type
defects in single layer silicene by analyzing the vibrational spectrum.
Although total energy calculations show the stability of the SW defects, our
phonon calculations reveals that there are two imaginary eigenfrequencies in
the SW-defected silicene spectrum (see Fig. \ref{fig-sw-phonon}).
For a 4$\times$4 supercell of silicene these modes
are located at $-14$ and $-20$~cm$^{-1}$. Analysis of the lattice dynamics of
structure for these two modes shows that the creation of a SW defect
results in large ripples around the defected region. As seen from Fig.
\ref{fig-sw-phonon}, the modes I and II imply the formation of sinelike and
cosinelike lattice distortions with respect to the midpoint of the
Si-Si dimer. Therefore the buckled planar structure of SW-defected silicene
is a local minimum on the potential energy surface. However, it is
worth to note that the stabilization of planar silicene structure by a
supporting surface has already been reported by several
groups.\cite{vogt,Feng,Fleurence} The existence of sinelike
and cosinelike rippling in silicene is in good agreement with the results for
SW-defected graphene.\cite{sw3} Although our phonon calculations revealed
the instability of densely SW-defected silicene, the results also imply the stability
provided by the long wavelength ripples. Therefore it is reasonable to assume that
the reactivity of defect cores for planar and long-wavelength-rippled silicene
do not differ significantly. Additionally, we perform molecular dynamics
calculations in order to further examine the thermal stability
of SW-defected silicene. At the temperature of 500 K, we choose time steps of 1
and also 2 fs. The SW defect and the surrounding bonds in silicene remain stable
and there is no indication of defect-healing throughout the molecular dynamics
simulations with duration of 2 ps at 500 K.

\begin{figure}
	\includegraphics[width=8.5cm]{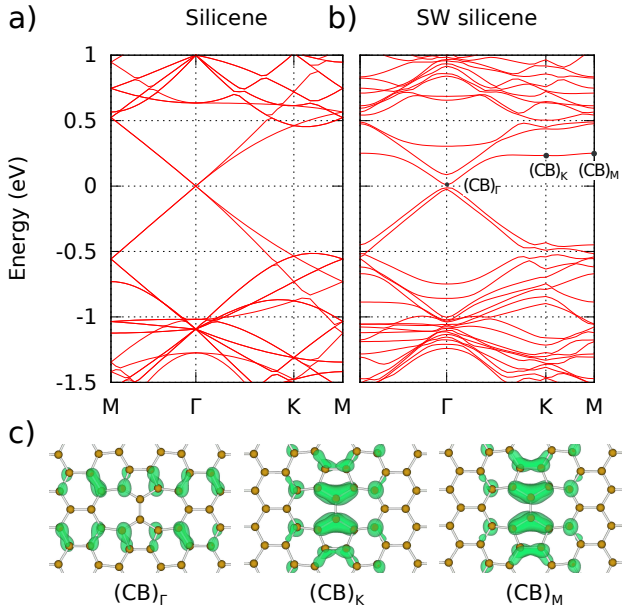}
	\caption{\label{fig-silicene_and_sw-el}
	(Color online) The electronic band structure for a 6$\times$6 supercell
	of (a) perfect and (b) SW-defected silicene. (c) Band decomposed
	charge densities of the conduction band around the Fermi level at
	$\Gamma$, K and M symmetry points (the isosurface is set to
	$0.57 \cdot	10^{-3}$ e/\AA$^{3}$)
	}

\end{figure}

In Figs. \ref{fig-silicene_and_sw-el}(a) and (b) the electronic band dispersion
of perfect and SW-defected silicene are presented. It is worth to note that
for a 6$\times$6 supercell, the K and K$^{\prime}$ symmetry points of the unit
cell of silicene are folded onto the $\Gamma$ point. Therefore, both valence
band maximums are transfered to the $\Gamma$ point of the supercell.
Clearly, as seen in Fig. \ref{fig-silicene_and_sw}(b), after the formation of a
SW defect the lattice symmetry of silicene is broken. Since the formation of
the SW defect breaks the six-fold symmetry of the silicene lattice, that results
in linearly crossing bands and the existence of highly mobile fermions in the
vicinity of the Dirac point, a band gap opening occurs at the crossing
point.\cite{hasan-mesh} The band gap opening for SW-defected silicene in a
6$\times$6 supercell is calculated to be 0.02~eV. It correspond to the system
with a defect concentration of 1.9$\cdot$10$^{13}$ cm$^{-2}$. The actual value
of the band gap depends on the concentration of the SW defects. In the case of a
5$\times$5 supercell with a single SW defect (2.7$\cdot$10$^{13}$ cm$^{-2}$) the
energy band gap increases to 0.1~eV. Here it is also worth to note that the
choice of a hexagonal supercell results in band edges that appear at the K
symmetry point ($\Gamma$ for 3$n\times$3$n$ supercells) and a temporarily
ordered defect configuration may result in a different band dispersion. Since
there are no dangling bonds introduced into the silicene lattice with the
creation of a SW defect, all the atomic orbitals of the Si atoms at the vicinity
of the defect are paired and hence there is no defect-originated magnetism.

\subsection{N-doped Silicene: Effect of Stone-Wales Defects}

\begin{figure}
\includegraphics[width=8.5cm]{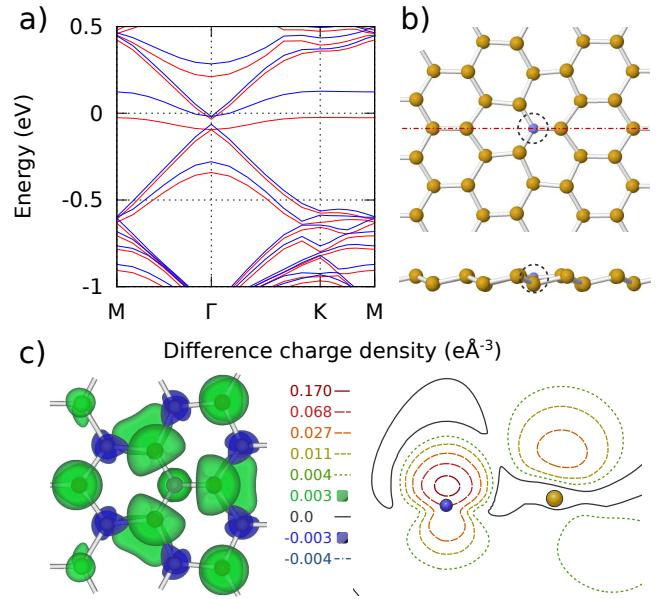}
\caption{\label{fig-n_sub}
(Color online) (a) The electronic band structure of N-doped silicene, (b) top
and side view of the structure. (c) 3D and contour plot of difference charge
density ($\rho_{\uparrow}-\rho_{\downarrow}$). The slicing plane is marked by
dot-dashed line in (b).}
\end{figure}

Doping materials with foreign atoms is an efficient way to manipulate their
electronic, magnetic and chemical properties. For carbon-based materials such
as graphite, graphene and nanotubes the doping especially with Group III or V
atoms is desirable since they induce $n$ or $p$ type doping and the
density of the charge carriers can be tuned by the concentration of the
substituents. It has been demonstrated that synthesis of N doped graphene
samples can be achieved by annealing in the presence of pyridine or NH$_{3}$
molecules and by chemical vapor deposition (CVD) and the observed features of
N-doped graphene are quite desirable for device applications.
\cite{kattel,Panchokarla,Zhao,Wang,Wei}
The formation energy of N substitution in graphene is 0.32 eV per N
atom.\cite{fujimoto_2011} Since N atom in graphene forms three
$\sigma$ bonds and $p_{z}$ orbital is filled by two electrons, N-doped graphene
shows nonmagnetic behavior.\cite{kattel} 
The N doping induces $n$-type doping of graphene and
shifts the Fermi level up by $\sim$ 0.6 eV for the case of 2 \% N
concentration.\cite{Panchokarla}

The substitutional doping of silicene with a N atom is of a different nature
as compared to the doping of graphene. The N-Si bond length in substituted
silicene (1.83 \AA) is significantly shorter than the Si-Si bond length in
pristine silicene (2.28 \AA) and the shorter bonds induce a local in plane
shrink deformation in the structure as presented in Fig. \ref{fig-n_sub}(b). The
nitrogen bonds with the neighboring Si atoms are in the plane, which reduces the
amount of local buckling. However, perfect silicene's lattice structure is
restored just several bonds away from the substituent atom. The formation energy
of N-doping in silicene is defined as

\begin{equation}
	E_{f} = E_{Tot} - (E_{Silicene} - \mu_{Si}) - \mu_{N}
\end{equation}

where $E_{tot}$ is the total energy of the N-doped silicene and $E_{Silicene}$
is the total energy of perfect silicene, $\mu_{Si}$ is the chemical potential
of a single silicon atom (calculated from the total energy of monolayer
silicene), and $\mu_{N}$ is the chemical potential of a nitrogen atom, defined
as one-half of the total energy of the $N_{2}$ molecule in the gas phase. This
choice of
chemical potentials is made to account for the stability of N-doped silicene
against $N_{2}$ molecular desorption. The binding energy of N-doped silicene
is here defined as:
\begin{equation}
	E_{bind} = E_{Tot} - (E_{Silicene} (n_{Si} - 1) / n_{Si}) - E_{N}
\end{equation}
where $n_{Si}$ is the number of Si atoms in a supercell of pristine silicene and
$E_{N}$ is the total energy of an isolated N atom.
The binding energy of a N atom in N-doped silicene is $-4.60$~eV. On the other hand, the
calculated formation energy is 0.59 eV as a consequence of the
large bond-dissociation energy of the $N_{2}$ molecule.

From the electronic band structure of N-doped silicene,
displayed in Fig. \ref{fig-n_sub}(a), it is seen
that the substitutional N atom turns semimetallic silicene into a ferromagnetic metal
with total magnetic moment of 1.0 $\mu _{B}$. The magnetic character of N
substituted silicene can be seen from the 3D and the contour
plot of the difference charge density ($\rho_{\uparrow}-\rho_{\downarrow}$)
in Fig. \ref{fig-n_sub}(c). It appears that the N-originated
state is delocalized over the nearest Si atoms and therefore N substitution
results in long-range spin polarization of the surrounding sp$^{3}$-like
orbitals even in a large 6$\times$6 supercell.
The magnetic behavior of N-doped silicene and delocalization of the
N-originated states are consistent with previously-reported results obtained
with the use of LDA functionals.\cite{silicene-Hasan,silicene-Jozef} 

\begin{figure}
	\includegraphics[width=7cm]{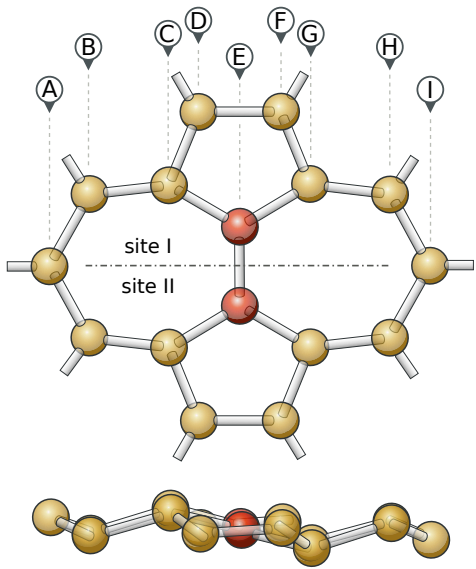}
	\caption{\label{fig-sw_defect}
	(Color online) The substitutional sites of the SW defect in silicene with
	associated letter, and the side view of the same structure.}
\end{figure}

Next we investigate the effect of the presence of a SW-defect on
substitutional nitrogen doping in silicene. As a result of local deformations
induced by the creation of a SW-defect (Fig. \ref{fig-silicene_and_sw}(b)) every
single Si atom in the surroundings of a SW-defect provides a unique place for
substitution. Since the surrounding atoms still maintain their buckled hexagonal
nature we will focus only on the Si atoms that are part of the heptagons or
pentagons in the SW-defect. As a consequence of the lattice symmetry with
respect to the Si-Si dimer at the defect core a similar behavior of adsorbates
on site I and II can be expected. However, in order to not exclude the possible
effects of lattice distortion on N substitution, calculations were performed for
all 16 sites of the SW defect. Possible positions for substitutional doping with
one substituent atom are displayed in Fig. \ref{fig-sw_defect}. Additionally, we
present structural properties, total magnetic moment, formation energies and
binding energy values in Table \ref{table-doping}.

The substitution positions can be divided into the four groups. The first
group contains the lattice sites that belong to edges of heptagons A, B, H and
I. We found that Si-N bonds at the edges of heptagons are not stable and hence
the substitution of a N atom on A, B, H and I positions cannot be realized.

\begin{table}
	\caption{\label{table-doping}
	Calculated values for N-doped on SW-defected 6$\times$6 silicene; bond
	length of N-Si ($d_{N-Si}$), angle between two of the lattice vectors
	of supercell of N-doped structure ($\alpha$),
	total magnetic moment of the system ($\protect\mu_{Tot}$),
	formation energy ($E_{f}$) and binding energy ($E_{bind}$).}

\begin{tabular}{lcccccc}
	\hline\hline	
	defect site & $d_{N-Si}$ & $\alpha$ & $\mu _{Tot}$ & $E_{f}$ & $E_{bind}$ \\
      & (\AA) & (deg) &$(\mu _{B})$ & (eV)& (eV) \\	
	\hline	
	\textbf{N @ pure silicene}  &1.83 1.83 1.83 & 60.0 &1.0&0.59&-4.6\\
	\textbf{N @ A,B,H,I} & - - -         &  -    & - & -  & - \\
	\textbf{N @ C}       &1.86 1.82 1.80 & 61.7  &1.0&1.8 &-3.4\\
	\textbf{N @ D}       &1.78 1.82 1.80 & 61.5  &0.7&1.1 &-4.0\\
	\textbf{N @ E}       &1.83 1.83 1.76 & 61.3  &1.0&1.5 &-3.7\\
	\textbf{N @ F}       &1.80 1.78 1.82 & 61.5  &0.7&1.1 &-4.0\\
	\textbf{N @ G}       &1.79 1.82 1.83 & 61.7  &1.0&1.8 &-3.3\\	
	\hline\hline
\end{tabular}

\end{table}

The second group is composed of C, and G substitution sites that locate at the
outermost intersection of heptagons and the pentagons of the defect core. Though
C and G sites are geometrically and energetically similar, because of the
lattice distortion induced by N-doping and buckled nature of the silicene
lattice, these adsorption sites are not identical. Similar to doping in perfect
silicene the substitution of a N atom on C and G site result in spin
polarization of 1 $\mu_{B}$. The binding energies on these lattice sites are
smaller compared to intrinsic silicene. 

The third group of possible substitution points are I and II sides of E which
are located in the middle of the defect core. Clearly, both sides of E are equal
for N-doping. When a N atom is substituted over E site, the resulting structure
has a total magnetic moment of 1 $\mu_{B}$ and doping on these sites is by
0.4~eV more favorable than doping on C and G sites.

The last group includes D and F substitution sites, which belong to the
pentagons of the SW-defect. The Si-N bond lengths in these lattice sites are
smaller or equal to the bond lengths in N-doped silicene. For those structures
we also observe large rippling of the order of 2~\AA. We found that
substitutional doping on D and F positions does not give rise to a shrinkage of
the lattice parameters. D and F points are energetically most favorable sites on
a SW defect with a binding energy of $-4.0$~eV. It is seen that nitrogen
substitution results in a net magnetic moment of 0.7 $\mu_{B}$. Similar to
N-doping in perfect silicene, SW-defected silicene shows metallic behavior for
all possible substitution sites that belong to the core of SW defect.

\section{Conclusions} \label{conclusion}

In this study we performed \textit{ab initio} calculations in order
to investigate the formation of SW defects in silicene and to show the effect of
these defects on the reactivity of the silicene with respect to nitrogen doping.
We found the formation of a SW-defect to be much easier in the buckled lattice
structure of silicene as compared to graphene. The softer bonding nature of
silicene allows easier formation of a SW defect with a smaller energy barrier
for both free standing as well as supported silicene. The presence of a SW
defect in silicene breaks the symmetry and results in a band gap opening in
electronic band structure with size depending on the defect concentration.
Additionally, our vibrational analysis reveals that the formation of SW defects
in freestanding silicene inevitably leads to the formation of large ripples.
Moreover, the presence of an underlying Ag(111) supporting surface increases the
barrier for the formation of SW defects from 2.4 to 2.8 eV. We also found that
the buckled nature of silicene provides a large energy barrier for healing of SW
defect and therefore defective silicene is quite stable even at high
temperatures. 

Furthermore, the presence of SW defects significantly modifies the doping
characteristics of silicene. While doping defect-free silicene by N atoms is
favorable with a little lattice shrinking, the presence of a SW defect limits
the number of possible doping sites. We found that all the nitrogen substitution
sites on the defect core are less preferable than defect-free silicene. Among
the possible sites of defect the edges of heptagons are least favorable sites
while doping of the N atoms at the edges of pentagons is most preferable. Our
findings on the reactivity of SW-defected silicene domains
agree well with the very recent study that reports the chlorine adsorption on
SW-defected graphene and carbon nanotube.\cite{mari} We believe that our results
provide a basis for the understanding of the characteristic properties of
defected silicene which are essential for its utilization in future electronics.

\begin{acknowledgments}
This work was supported by the Flemish Science Foundation (FWO-Vl) and the
Methusalem foundation of the Flemish government. Computational resources were
provided by TUBITAK ULAKBIM, High Performance and Grid Computing Center (TR-Grid
e-Infrastructure), and HPC infrastructure of the University of Antwerp (CalcUA)
a division of the Flemish Supercomputer Center (VSC), which is funded by the
Hercules foundation. H.S. is supported by a FWO Pegasus Marie Curie Fellowship.
\end{acknowledgments}

\end{document}